\documentclass[aps,pra,twocolumn]{revtex4-2}
\usepackage{amsmath,epsfig,mathrsfs,graphicx,amssymb,color,float}
\usepackage[T1]{fontenc}
\usepackage{t1enc}
\usepackage{hyperref}
\usepackage{array}
\usepackage{booktabs}
\usepackage{color}

\def\be#1\ee{\begin{equation}#1\end{equation}}
\newcommand{\ba}{\begin{eqnarray} }
\newcommand{\ea}{\end{eqnarray} }

\usepackage[cp1250]{inputenc}   
\usepackage{amsmath}            
\usepackage{amsfonts}           
\usepackage{amssymb}            
\usepackage{enumerate,graphicx}
\usepackage{array}
\usepackage{booktabs}
\usepackage{yquant}
\usetikzlibrary{fit, quotes}
\usetikzlibrary{snakes,arrows,shapes}
\useyquantlanguage{groups}

\usepackage{physics}

\def\mb{\begin{pmatrix}}
\def\me{\end{pmatrix}}
\def\be#1\ee{\begin{equation}#1\end{equation}}

\begin{document}
                
\title{Closing the objectivity loophole in Bell tests on a public quantum computer}

\author{Adam Bednorz$^1$}
\email{Adam.Bednorz@fuw.edu.pl}
\author{Josep Batle$^{2,3}$}
\author{Tomasz Bia{\l}ecki$^{1,4}$}
\author{Jaros{\l}aw K. Korbicz$^5$}
\email{jkorbicz@cft.edu.pl}

\affiliation{$^1$Faculty of Physics, University of Warsaw, ul. Pasteura 5, PL02-093 Warsaw, Poland}
\affiliation{$^2$Departament de F\'isica and Institut d'Aplicacions Computacionals de Codi Comunitari (IAC3), Campus UIB, E-07122 Palma de Mallorca, Balearic Islands, Spain}
\affiliation{$^3$CRISP -- Centre de Recerca Independent de sa Pobla, 07420 sa Pobla, Balearic Islands, Spain}
\affiliation{$^4$Faculty of Physics and Applied Informatics, University of Lodz,
ul. Pomorska 149/153, PL90-236 Lodz, Poland}
\affiliation{$^5$Center for Theoretical Physics, Polish Academy of Sciences, al. Lotnik{\'o}w 32/46, PL02-668 Warsaw, Poland
}

\begin{abstract}
We have constructed and run a Bell test of local realism focusing on the objectivity criterion.
Objectivity means that the outcomes are confirmed macroscopically by a few observers at each party.
The IBM Quantum and IonQ devices turn out to be sufficiently accurate to pass such an extended Bell-type test, although at the price
of communication loopholes and residual but statistically significant signaling. The test also serves as the benchmark of 
entanglement spread across larger sets of qubits.
\end{abstract}

\maketitle

\section{Introduction}

Quantum mechanics is incompatible with local realism, which is confirmed by violation of Bell inequalities by two separate parties \cite{epr,bell,chsh}.
The violation relies on several assumptions to be satisfied experimentally.
First of all, the detection must be complete; every event must be recorded and taken into account.
In the past, due to low efficiency of photodetectors, most photons got lost, so one applied postselection, 
reduction of outcomes to the cases where each photon has clicked \cite{belx1a,belx1b,belx1c,belx1g}.
It left the detection loophole open, since some local realistic models could exploit low detection rate \cite{game,loop1,loop2,larsson}.
In modern experiments, photodetection efficiency has increased sufficiently to abandon postselection.
Other systems, ions, atoms and transmons are always fully detected, so this loophole has never affected them \cite{belx1d,belx1h,belx1e,belx1f}.
Another challenge is locality, meaning that the two parties are sufficiently separated so that no exchange of information
is possible during a single experiment. When the parties were close to each other, which was the case of
atoms, ions, and transmons, the timescale was long enough when compared to the distances, and thus communication was in principle possible.
To prevent possible communication, one can make a large separation, of the order of tens of meters and more, and make the experiment quick enough
\cite{hensen,nist,vien,munch,storz}.
The speed of light is taken as the absolute limit of communication, although  it is an axiom \cite{wight} and current Bell tests are done in some preferred frame, where 
relativistic invariance is not assumed. One should rather eliminate potential communication channels, like cables and connections, making locality a more contextual than fundamental assumption. The communication loophole is connected with the freedom of choice because the signal cannot be predetermined. Otherwise,
there is no information to reveal to the other party, regardless of the spacetime considerations. Bell tests assume randomly chosen measurement settings, which must occur locally
so that information about the choice of one party does not reach the other before the measurements are accomplished.
If one insists on the lack of such freedom, claiming superdeterminism -- predetermination of all measurement settings in the past, the communication, although
very conspiratorial, cannot be excluded. One has to refer to common sense that the choices are truly and securely random at some stage, 
by the technical constraints of the setup. Otherwise, no conclusion can be drawn, not only from Bell tests, but all experiments relying on
some external influence independence. The other side of the communication problem is the end of the measurement. While it is tempting to
mark the accomplishment of the readout by a control clock, it is again a matter of common sense agreement. 
It is better exposed by the Wigner friend problem, where the measurement is performed by an agent (friend) 
which is observed by another apparatus or human (Wigner) \cite{wigf,wigf1,wigf3,wigf4}. The end of the measurement is marked by the moment when all or the most of the external observers receive the outcome.
Just like superdeterminism, it becomes a matter of convention or common sense, referring to the technical specification of the detector.
The end of the measurement is the moment when the readout outcome becomes objective, visible to many independent observers. One can define quantum objectivity
more rigorously by introducing the quantum Darwinism idea \cite{Zurek_book_2025} and  Spectrum Broadcast Structures (SBS) \cite{object,object2, korb2021}, which have been already applied to the analysis of quantum measurements \cite{korb2017}. 

Motivated by the objectivity research, in the hereby paper, we rerun Bell tests on IBMQ and IonQ quantum platforms, simulating 3 observers (friends) at each side, with the requirement that the result $+1$ or $-1$ must be
unanimous among the party to be assigned, otherwise we set $0$. It builds on the previous tests with 2 friends \cite{friendslo,friendslo1,friendslo2}
by more friends, more statistics, no postselection, and much more extensive analysis, including no-signaling.
It is different from the test of local friendliness \cite{wigf1}, where the additional observers could make measurements \emph{instead} of the main ones,
but not simultaneously, postselection on coincidences was applied, and the measurement sets were predetermined (no freedom of choice).
Local friendliness assumes absoluteness of events, i.e. that observers have access to the same values, while we check it without assuming.
 The test has been run on IBM Quantum, which is a routine platform for
quantum experiments nowadays \cite{ibm,qis,ibmedu,ibm-bell,ibmb,alsina,garcia}, including tests of objectivity \cite{objwit}. The light-crossing time is far shorter than the gate and readout times
in  IBM devices to close the communication loophole. Nevertheless, we have minimized the possibility of communication by (i) separating the parties by a middle qubit, and (ii)
applying only phase-dependent settings to avoid crosstalk in the energy eigenbasis. The choice of the setting was pseudorandom, i.e. the angles were randomly shuffled, 
so that only conspiratorial access of the detector to the initial data could affect freedom of choice. Together with the Bell test, we also checked no-signaling in the context of
IBM Quantum and IonQ limitations. No-signaling means that the setting of one party cannot affect the readout of the other one. In contrast to Bell inequality,
when experiments are performed respecting space-like separation, nonsignaling is respected both in classical and quantum theory \cite{hall}. At the fundamental level, signaling would exceed the relativistic information transfer speed limit, namely the speed of light. We stress that all performed Bell tests do not directly test relativity as they are set in a preferred frame, and the actual signal speeds through fibers or  waveguides are usually smaller than the vacuum limit. Violation of no-signaling indicates at least some unspecified crosstalk, and poses a serious technical
hurdle if its causes cannot be identified. In the test with friends, one has to check all 8 combinations of outcomes.

The results show violation of Bell inequality in the stringent version with unanimity condition, but also some violation of no-signaling.
The latter is not large in the sense of absolute values, compared to Bell, but  still significant, beyond 10 standard deviations. 
The test is a simple benchmark of multipartite entanglement as its success relies on maintaining entanglement among friends.

The paper is organized as follows. We start with a brief recapitulation of the Bell test, extended to multifriend parties, and definition of no-signaling.
Then we explain the actual implementation on IBM Quantum and IonQ. Next, we show and comment the results, and conclude the paper with a discussion of interpretation and possible next steps and challenges.

\section{Bell test with friends and objectivity} 

The standard Bell test involves two parties, $A$ and $B$, who share an entangled state and make local measurements independent of each other.
For consistency, we shall work in the qubit basis $|0\rangle$ and $|1\rangle$ for each party and assume the entangled state $|\psi\rangle$ of the form
\be
\sqrt{2}|\psi\rangle=|00\rangle-i|11\rangle
\ee
with the local measurement operators 
\be\label{pomiar}
A=ie^{i\alpha}|1\rangle\langle 0|-ie^{-i\alpha}|0\rangle\langle 1|
\ee
and analogously $B$ with $\alpha\to\beta$. The projective measurements in the eigenbasis of $AB$ give outcomes $A,B=\pm 1$ and the correlation
\be
\langle AB\rangle=-\sin(\alpha+\beta).
\ee
Let each party choose its setting by taking $\alpha_a$, $\beta_b$ for $a,b=0,1$, i.e. each party has two independent measurement choices.
We shall denote $A_a$ and $B_b$ for these respective choices.
For all classical joint probability distributions $P(A_0,A_1,B_0,B_1)\geq 0$, $\sum P=1$
the Clauser-Horne-Shimony-Holt (CHSH) \cite{chsh} inequality holds
\be
\mathcal B=\langle A_0B_0\rangle-\langle A_1B_0\rangle-\langle A_0B_1\rangle-\langle A_1B_1\rangle\leq 2\label{chshi}
\ee
with correlations defined as $\langle F\rangle=\sum_{\Omega}P(\Omega)F(\Omega)$.
However, in quantum mechanics, setting the  angles $\alpha_{0,1}=0,\pi/2$ and $\beta_{0,1}=-\pi/4,+\pi/4$ we get
\be
\langle A_1B_1\rangle=\langle A_1B_0\rangle=\langle A_0B_1\rangle=-\langle A_0B_0\rangle=-1/\sqrt{2}
\ee
which  violates (\ref{chshi}) at $2\sqrt{2}\simeq 2.828$. It means that the joint probability cannot exist, and
 the outcomes of the experiment $A_0$, $A_1$, $B_0$, $B_1$ are not a set of counterfactually definite values.

\subsection{Objectivity loophole and how to close it}

The catch in the original test is the measurement process, hidden in the projection postulate. In reality, no measurement is perfect, and moreover,
it needs some criterion that it has been completed. 
Suppose an unspecified sub-luminal interaction (or an adversary in a device-independent scenario) could, in principle, change the observer's registered value after the readout time tag based on the information about the other party's input choice. Then, the violation of the CHSH inequality would be explainable by a local hidden variable model supplemented with communication since the final time tag would already be within the lightcone.

A solution might be to simply copy the readout of the measurement to an independent classical memory along with the final time tag and store both copies.
Then again an adversary could  try to change the result, although harder. There is no a priori number of copies to justify the objective conclusion.
A reasonable claim is that a measurement is completed if the result is objective, accessible to many observers. This requires that the measurement result is stored faithfully in many copies in the macroscopic apparatus and those copies can be accessed by external observers. As an example, one may think of a single pixel of the measuring device, which by emitting or reflecting light can be seen by many observers reading out the measurement result.

This loophole is complementary to the freedom of choice occurring at the beginning of the test, as it concerns the finish.
The classical copying is more vulnerable to attacks than a quantum one as it is done after the quantum measurement, when the coherence is already lost.
By copying at the quantum stage, the adversary can be detected by reversal of the measurement and changing the settings as in local friendliness scenario
\cite{friendslo2}.  In contrast to local friendliness, the identity of the values among friends is not assumed but directly verified.

\begin{figure}
\begin{tikzpicture}[scale=1]
		\begin{yquant*}[register/separation=3mm]
			init { $\ket 0$} q[0,1,2,3,4,5,6];
			box {$S_0$} q[3];
			cnot q[4] | q[3];
			cnot q[2] | q[3];
			cnot q[3] | q[2];
			align -;
			box {$S_\alpha$} q[2];
			box {$S_\beta$} q[4];
			cnot q[1] | q[2];
			cnot q[0] | q[2];
			cnot q[5] | q[4];
			cnot q[6] | q[4];
			
			measure q[0-2];
			measure q[4-6];
			output {$A^2$} q[0];
			output {$A^1$} q[1];
			output {$A^0$} q[2];
			output {$M$} q[3];
			output {$B^2$} q[6];
			output {$B^1$} q[5];
			output {$B^0$} q[4];
		\end{yquant*}
\end{tikzpicture}
\caption{The circuit of the Bell test with friends. The $CX$ gates are denoted by the line between control (depicted as $\bullet$) and
target (depicted as $\oplus$). The operational flow is from the left to the right.}\label{ciro}
\end{figure}
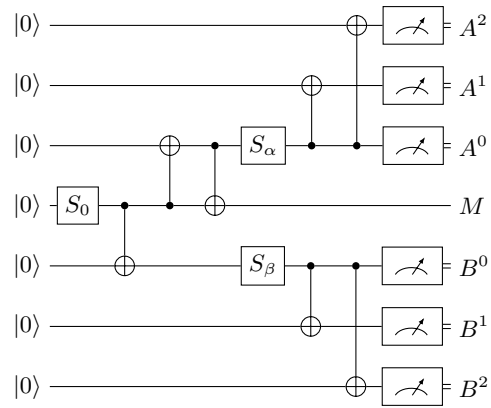

Thus, in the  friend(s)/objectivity scenario we are simulating here, information is copied to auxiliary systems before measurement.
This simulates the information proliferation in the degrees of freedom of the measuring apparatus.  Imagine three observers (friends)
 $A^0$, $A^1$ and $A^2$ with the joint
space of states $|A^0A^1A^2\rangle$  with  $A^i=0,1$ and analogously for $B$; see Figs. \ref{ciro} and \ref{objsch}. We will consider projective measurements on the three qubits giving 8 outcomes of the form $A^0A^1A^2$
corresponding to projection onto the respective basis states.
The initial entanglement is between qubits $A^0$ and $B^0$, i.e. the initial state is
\be
\sqrt{2}|\psi\rangle=|000,000\rangle-i|100,100\rangle
\ee
Now the choice of the Bell measurement \eqref{pomiar} is realized by a unitary rotation
\be
S_\alpha=\frac{1}{\sqrt{2}}\begin{pmatrix}
1&-ie^{i\alpha}\\
-ie^{-i\alpha}&1
\end{pmatrix}\label{saa}
\ee
acting on the qubit $A^0$ in the basis $|0\rangle$, $|1\rangle$ (and $B^0$ with $\beta$).
Unlike in the usual Bell scenario, where the state is kept fixed from run to run and the measurements are changed, here we use an equivalent approach of changing the local state from run to run but keeping the measurement fixed. This is dictated by the technical capabilities of the simulation platforms used and our desire to distribute as precisely the same information to all observers as possible. Therefore, it is better, error-wise, to set the measurement angles before the information copying rather than angle-dependent coupling individually each observer to the system. Since we want to test the Bell-type scenario, the rotation must take place before broadcasting information to the friends.  The other choice, i.e. individual basis rotation by each observer would lead to a different type of scenario with multipartite entanglement and appropriate generalization of Bell inequalities \cite{svetlichny,mermin}. At this stage the state reads
\begin{align}
&\sqrt{8}|\psi_{\alpha\beta}\rangle=(1+ie^{i(\alpha+\beta)})|000,000\rangle\nonumber\\
&-(i+e^{-i(\alpha+\beta)})|100,100\rangle\nonumber\\
&
-(ie^{-i\alpha}+e^{i\beta}) |100,000\rangle\nonumber\\
&-(ie^{-i\beta}+e^{i\alpha}) |000,100\rangle.
\end{align}
Then we apply the crucial step of information copying. We simulate it via conditional flipping of qubits $A^{1,2}$ if $A^0=1$, i.e.
we apply a controlled-$X$ gate of the form $CXX|A^0A^1A^2\rangle=|A^0(A^1\oplus A^0)(A^2\oplus A^0)\rangle$ with $F\oplus G=(F+G)$ modulo $2$; see Fig. \ref{ciro}. In the full basis $|A^0A^1A^2\rangle$, this operation reads
\be
\begin{pmatrix}
1&0&0&0&0&0&0&0\\
0&1&0&0&0&0&0&0\\
0&0&1&0&0&0&0&0\\
0&0&0&1&0&0&0&0\\
0&0&0&0&0&0&0&1\\
0&0&0&0&0&0&1&0\\
0&0&0&0&0&1&0&0\\
0&0&0&0&1&0&0&0
\end{pmatrix}.
\ee
The state afterwards reads:
\begin{align}\label{final}
&\sqrt{8}|\bar{\psi}_{\alpha\beta}\rangle=(1+ie^{i(\alpha+\beta)})|000,000\rangle\nonumber\\
&-(i+e^{-i(\alpha+\beta)})|111,111\rangle\nonumber\\
&-(ie^{-i\alpha}+e^{i\beta}) |111,000\rangle\nonumber\\
&-(ie^{-i\beta}+e^{i\alpha}) |000,111\rangle.
\end{align}
Strictly speaking, from the quantum measurement theory, this is not yet the final state of the measurement process, as it is obviously entangled between the results $0$ and $1$. This is the pre-measurement state, which forms after information has been encoded in many degrees of freedom of the measuring apparatus.  The final post-measurement state is obtained after tracing out some degrees of freedom as unobserved, which in the ideal case induces perfect decoherence and collapse of entanglement into a mixture. However, due to the limited connectivity of the used quantum devices, and hence a limited amount of qubits one can reliably include in the simulation of the measuring apparatus, we will not discard any of the qubits in \eqref{final} but rather perform the strict majority vote measurements on each side. We thus require unanimity between observers, i.e.
\begin{align}
&000\to +1,\nonumber\\
&111\to -1,\nonumber\\
&\mbox{ otherwise }\to 0\label{restr}
\end{align}
so the measured observable essentially reads
\be
\bar{A}=|000\rangle\langle 000|-|111\rangle\langle 111|
\ee
and similarly for $B$. In the ideal case 
\be
CXX\:\bar{A}\:CXX=|000\rangle\langle 000|-|100\rangle\langle 100|
\ee
so the correlations are identical to the Bell test with single observers (\ref{chshi}) for $A\to\bar{A}$ and $B\to\bar{B}$.
Moreover, only the outcomes $000$ and $111$ should be registered, but we do not make any postselection in correlations,
calculating the probability with respect to all possible outcome configurations.

Such a test shares the state before measurement.
Note that we can do it only after the setting rotation. We cannot clone the state before choosing and applying the $\alpha/\beta$-dependent rotation.
Otherwise, we would get a highly entangled state
\be
(|000,000\rangle-i|111,111\rangle/\sqrt{2},
\ee
which makes the friends unable to receive identical outcomes.
The individual probabilities are given by projective measurements
\be
P(A^0A^1A^2,B^0B^1B^2)=|\langle A^0A^1A^2,B^0B^1B^2|\psi_{\alpha\beta}\rangle|^2
\ee
for all values $A^j,B^j=0,1$, which gives
\begin{align}
&P(000,000)=P(111,111)=(1-\sin(\alpha+\beta))/4,\nonumber\\
&P(000,111)=P(111,000)=(1+\sin(\alpha+\beta))/4,\nonumber\\
&P(A^0A^1A^2,B^0B^1B^2)=0\mbox{ otherwise }
\end{align}
The violation of Bell inequality (\ref{chshi}) with the restriction (\ref{restr}) by the state (\ref{final}) is therefore essentially identical, but the result gets a new meaning. It is confirmed collegially by all friends who must be unanimous to count it.

\subsection{No-signaling}

In the test, one can simultaneously verify no-signaling, i.e., independence of the result of friends at one party of the setting chosen by the other one.
In particular, the probability  $P(AB|ab)$ for the settings $a$ and $b$ chosen by the team $A$ and $B$, respectively, gives
\begin{align} 
&\delta  P_{a\ast}(A)=P(A\ast|a0)-P(A\ast|a1)\nonumber\\
&\delta  P_{\ast b}(B)=P(\ast B|0b)-P(\ast B|1b)\label{nos}
\end{align}
with $\ast$ meaning the marginal probability, i.e. ignoring (summing up) the results of the other party.  By no-signaling principle $\delta P$ 
must remain $0$ within statistical error. In the ideal case 
\begin{align}
&P(000,\ast\ast\ast|a\ast)=P(111,\ast\ast\ast|a\ast)=1/2,\nonumber\\
&P(\ast\ast\ast, 000|\ast b)=P(\ast\ast\ast, 111|\ast b)=1/2,
\end{align}
while all other probabilities are $0$. 
In reality, imperfections make them not zero, but still no-signaling (\ref{nos}) should hold.
The deviations from equality (\ref{nos}) can be nonzero but should remain within the statistical error.
No-signaling can be tested very accurately in this case, but in practice only well space-like separated experiments 
are free from fundamentally allowed communication.

\section{Setup on IBM Quantum}

IBM Quantum provides a set of native gates for single and two-qubit operations.
For single-qubit rotation, we can essentially use (\ref{saa}), which is realized natively.

To create entanglement we have to first apply $S_0$ to a qubit and then use a two qubit coupling, $CX$ (also known as $CNOT$) gate: $CX_\downarrow|FG\rangle=|F(F\oplus G)\rangle$ gate reverses the state $F$ when $G=1$,  while $CX_\uparrow|FG\rangle=|(F\oplus G)G\rangle$ acts in the opposite direction, see details in Appendix \ref{appa}.
The pair of $CX$ gates swap the state with $|0\rangle$, i.e.
\be
CX_\uparrow CX_\downarrow|\phi 0\rangle=|0\phi\rangle
\ee
To create the entangled state we initialize an auxiliary qubit $M$, see Fig. \ref{objsch},  by $|0\rangle\to S_0|0\rangle=|-_y\rangle=(|0\rangle-i|1\rangle)/\sqrt{2}$, and apply $CX_\downarrow$ to $|-_y\; 0\rangle$ 
to get $
(|00\rangle-i|11\rangle)\sqrt{2}$ for qubits $M$ and $B^0$. The source $M$ qubit is then swapped to $A^0$ by the pair of opposite $CX$ gates.

The settings are defined by $S_\alpha$ and $S_\beta$ rotations applied to the final qubits $A_0$ and $B_0$, as described in the previous section. 
The state is then copied to the friends by a pair of $CX$ gates, see the full circuit in Fig. \ref{ciro}.

The statistical error can be evaluated for a single shot and scaled by the number of repetitions, as the accumulation
of  Bernoulli statistics (Gaussian at large number of repetitions).
A single shot gives the variance for correlations
\be
\langle\Delta^2 (AB)\rangle=\langle A^2B^2\rangle-\langle AB\rangle^2,
\ee
which is $1/2$ in the ideal case.
For the test of signaling, we take the simple Bernoulli formula 
\be
\langle(\Delta P)^2\rangle=P(1-P)
\ee
which is ideally $1/4$ for $000$ and $111$ while $0$ otherwise. In practice, the latter is not zero but still very small, which allows us to check no-signaling
at very small error.

\begin{figure}
\includegraphics[scale=.5]{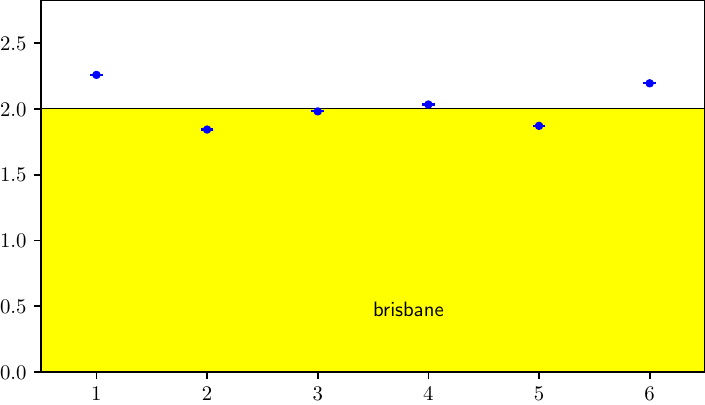}
\includegraphics[scale=.5]{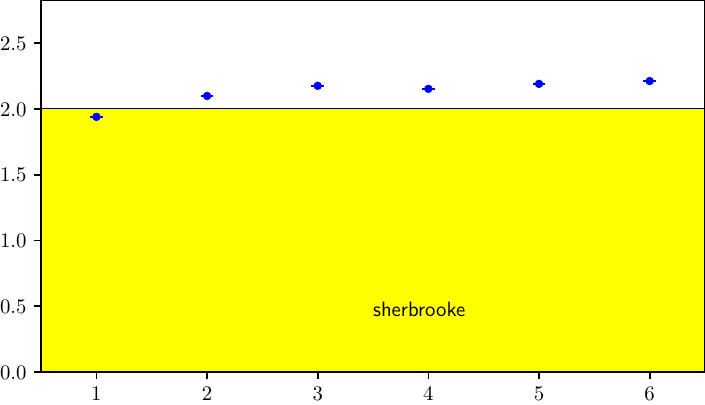}
\includegraphics[scale=.5]{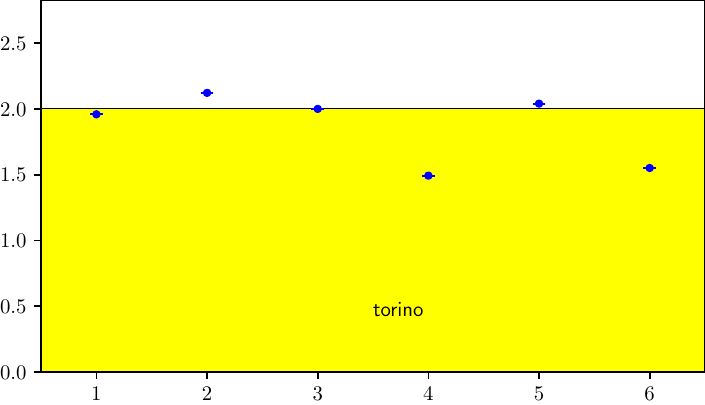}
\includegraphics[scale=.5]{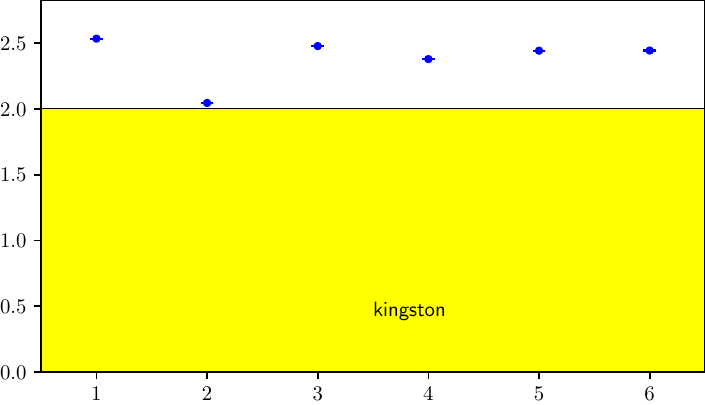}
\caption{Bell test results, the value $\mathcal B$ (\ref{chshi}),  with objectivity condition i.e. all 3 observers at each party must agree on the outcome on \emph{ibm\_brisbane},
\emph{ibm\_sherbrooke},  \emph{ibm\_torino},
\emph{ibm\_kingston}. The horizontal axis represents the index of the set of qubits. The classical region $\mathcal B\leq 2$ is yellow. The error bars are $1\sigma$.}
\label{belo}
\end{figure}

\section{Results}

\begin{figure}
\includegraphics[scale=.7]{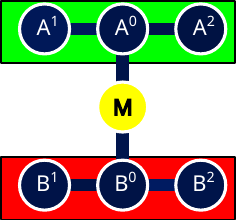}
\caption{The connections in the objectivity test. The middle (yellow) qubit $M$ lies between the two parties $A$ and $B$ (green and  red, respectively);
the main qubits $A^0$ and $B^0$ are coupled to the friends.}\label{objsch}
\end{figure}

\begin{figure*}
	\includegraphics[scale=.6]{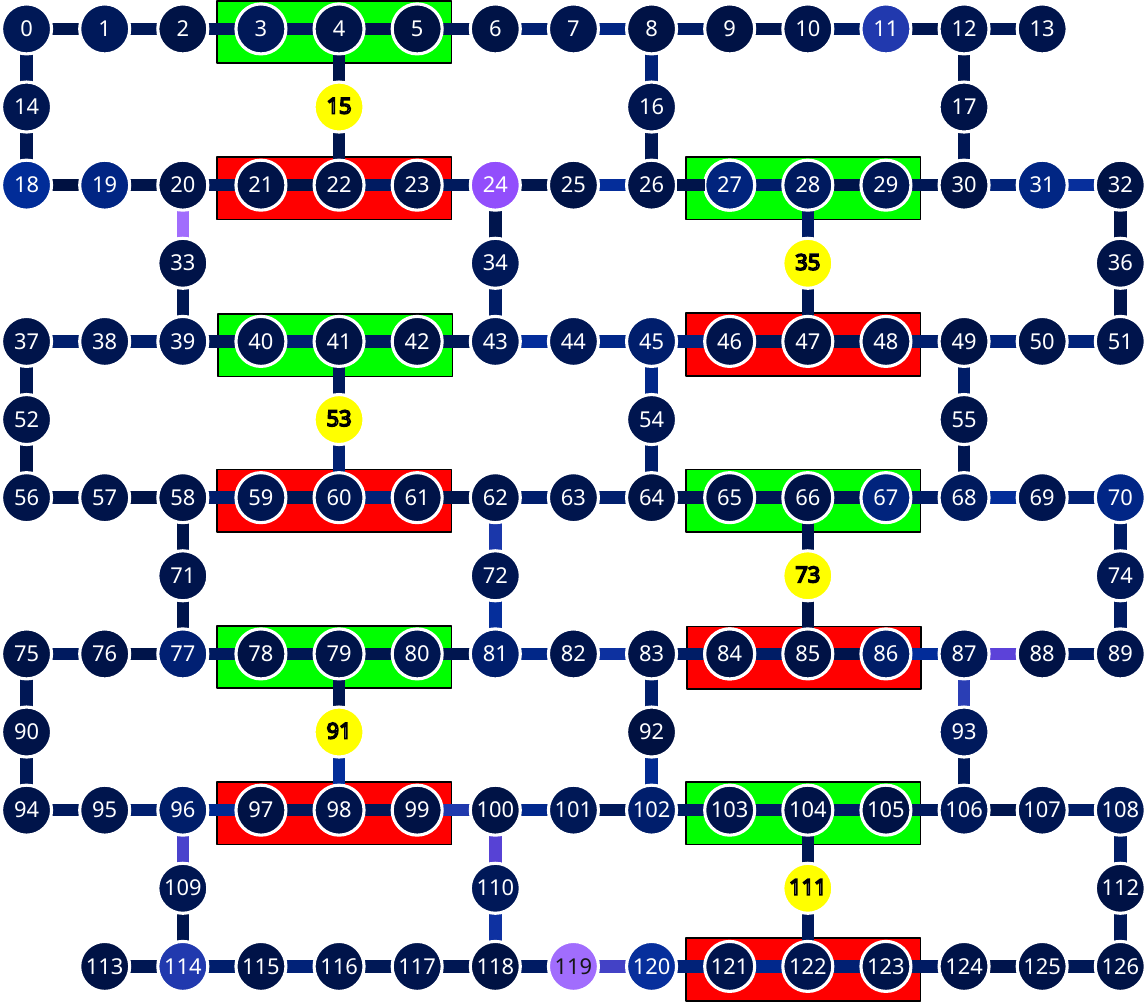}
\caption{The placement of all the tested groups on the \emph{ibm\_brisbane} and \emph{ibm\_sherbrooke},
denoted  as in Fig. \ref{objsch}. Adapted from IBM Quantum \cite{ibm}.}\label{objqub}
\end{figure*}

\begin{figure*}
	\includegraphics[scale=1]{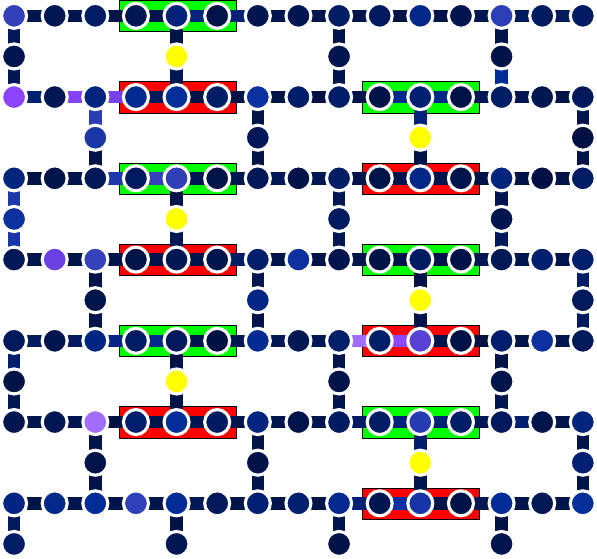}
\caption{The placement of all the tested groups on the \emph{ibm\_torino},
denoted  as in Fig. \ref{objsch}. Adapted from IBM Quantum \cite{ibm}.}\label{objqubt}
\end{figure*}

\begin{figure*}
	\includegraphics[scale=1]{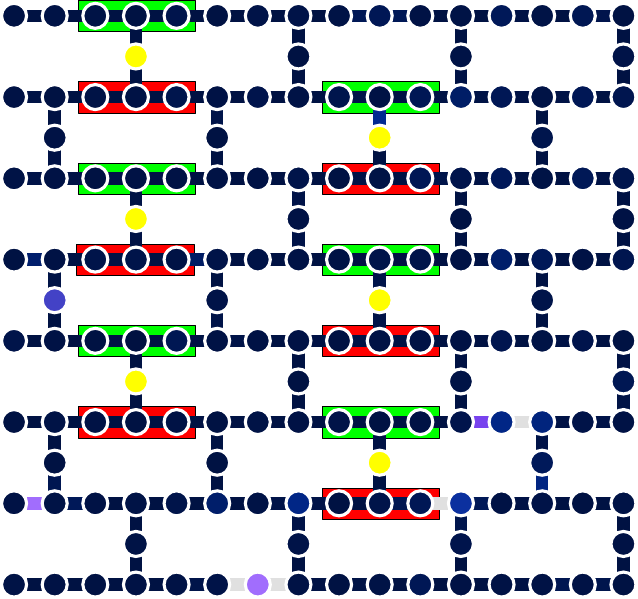}
\caption{The placement of all the tested groups on the \emph{ibm\_kingston},
denoted  as in Fig. \ref{objsch}. Adapted from IBM Quantum \cite{ibm}.}\label{objqubk}
\end{figure*}

\begin{figure*}
\includegraphics[scale=.5]{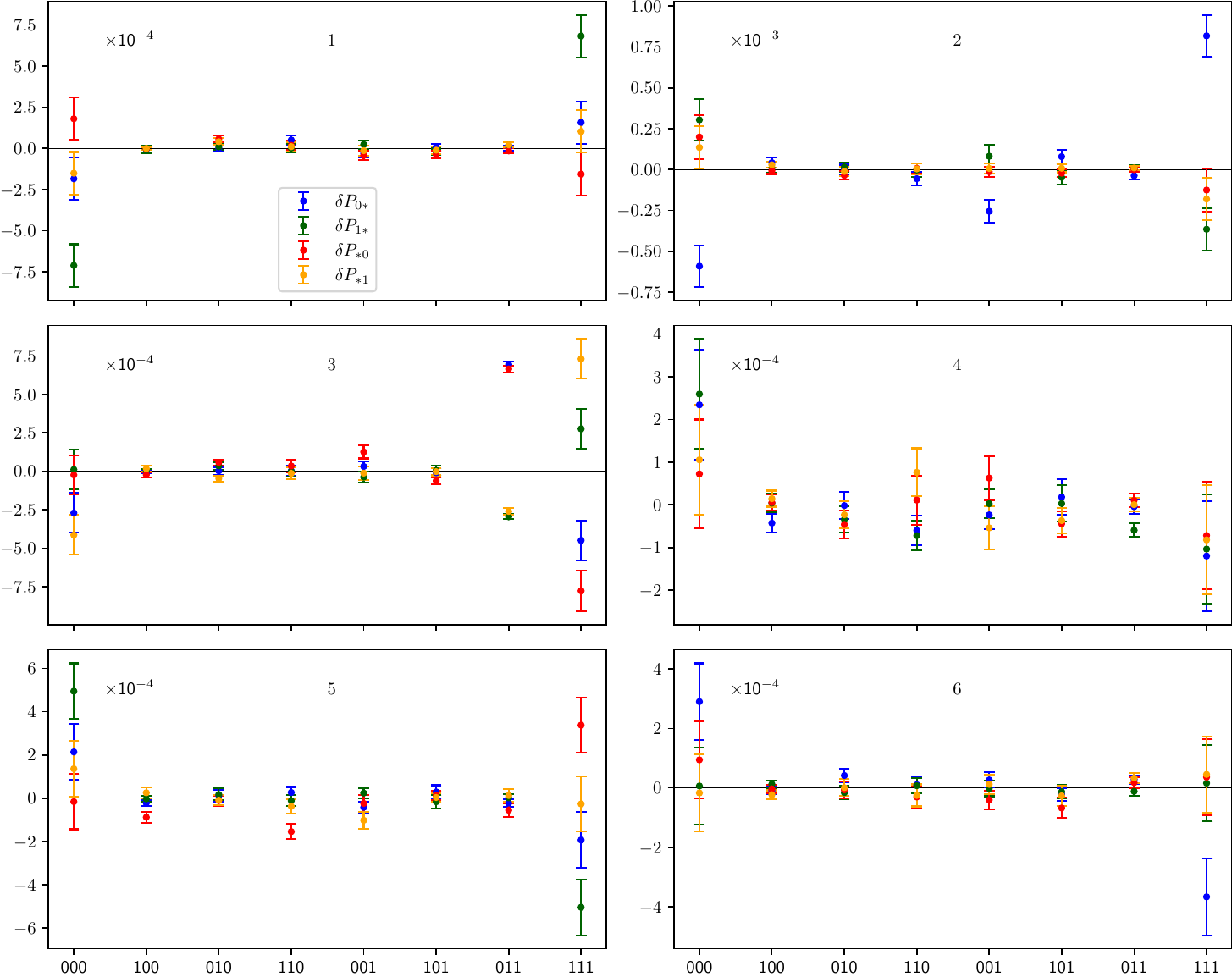}
\caption{Signaling between qubits $\delta P_{a\ast}(A^1A^2A^0)$ and $\delta P_{\ast b}(B^1B^2B^0)$  in each of the groups on \emph{ibm\_brisbane}. Notation as in Fig. \ref{belo}.}
\label{ss1b}
\end{figure*}

\begin{figure*}
\includegraphics[scale=.5]{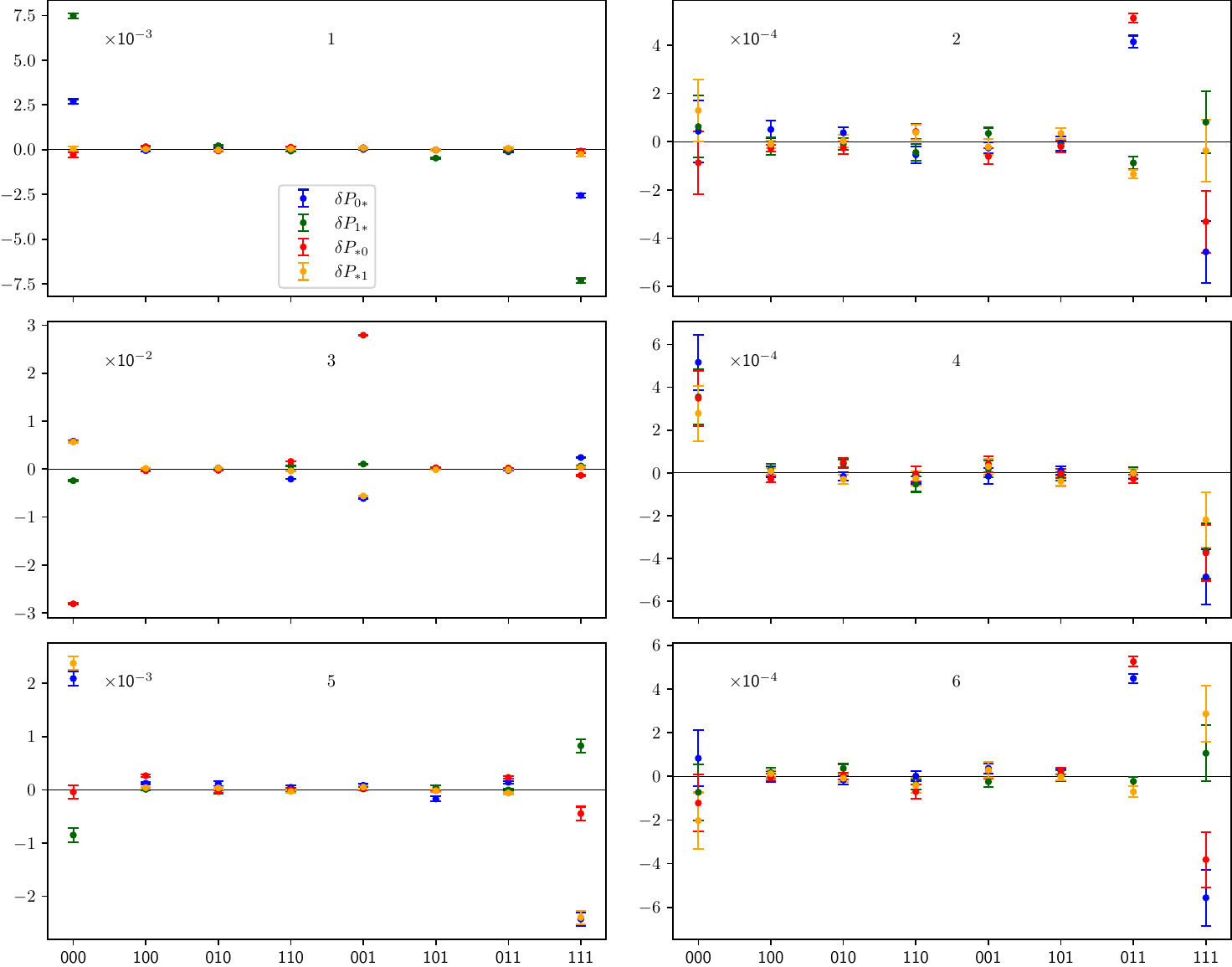}
\caption{Signaling between qubits $\delta P_{a\ast}(A^1A^2A^0)$ and $\delta P_{\ast b}(B^1B^2B^0)$  in each of the groups on \emph{ibm\_sherbrooke}. Notation as in Fig. \ref{belo}.}
\label{ss1s}
\end{figure*}

\begin{figure*}
\includegraphics[scale=.5]{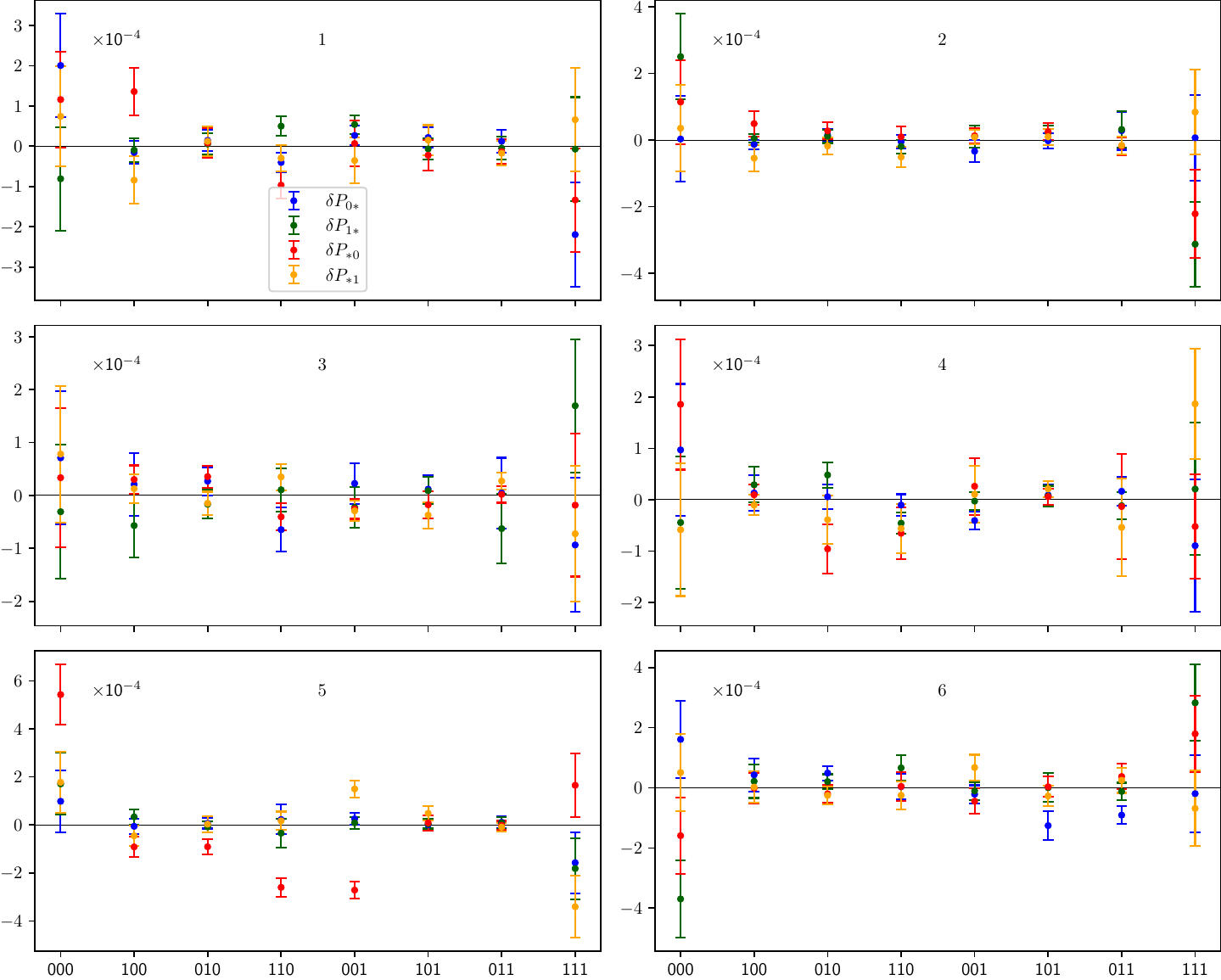}
\caption{Signaling between qubits $\delta P_{a\ast}(A^1A^2A^0)$ and $\delta P_{\ast b}(B^1B^2B^0)$  in each of the groups on \emph{ibm\_torino}. Notation as in Fig. \ref{belo}.}
\label{ss1t}
\end{figure*}

\begin{figure*}
\includegraphics[scale=.5]{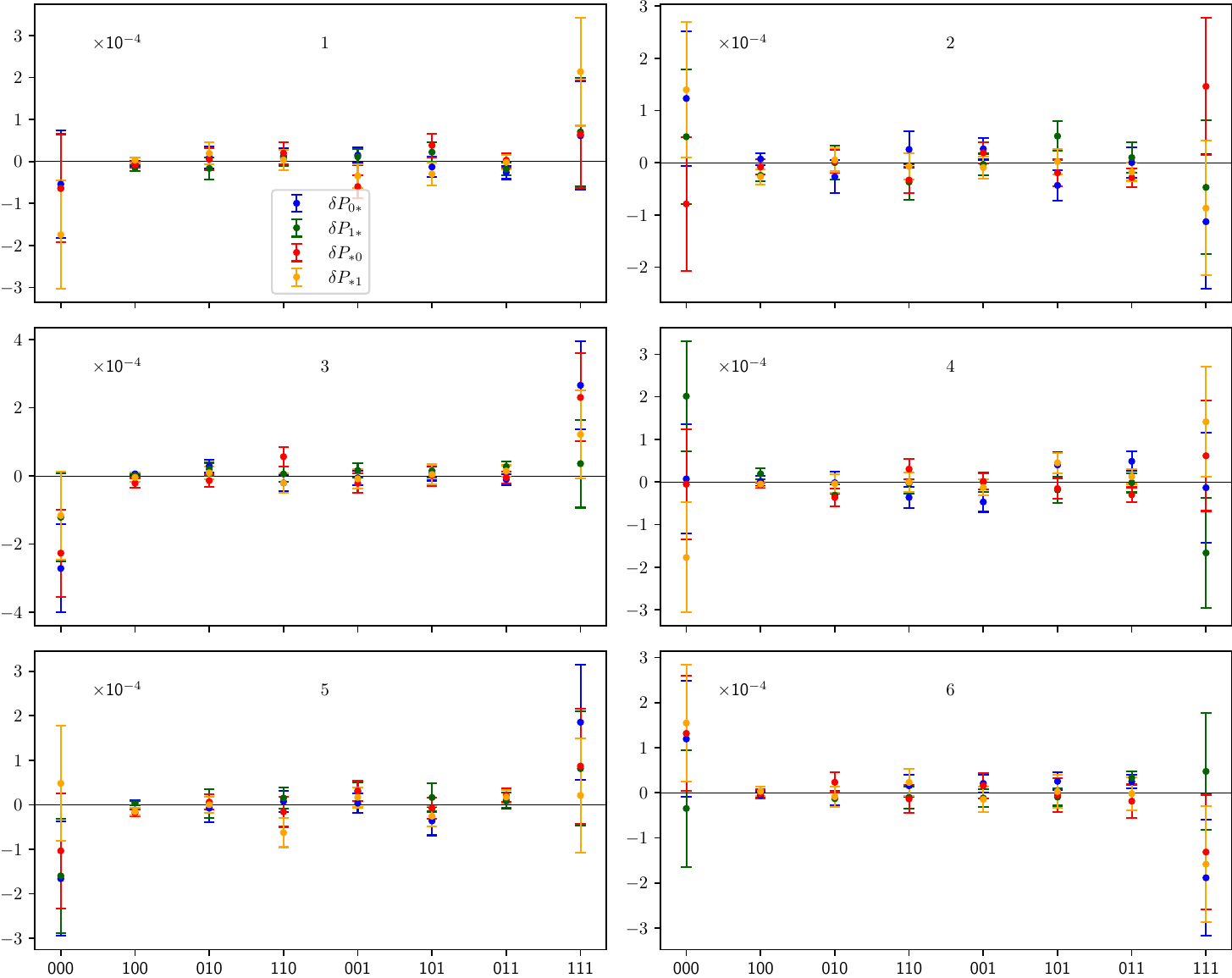}
\caption{Signaling between qubits $\delta P_{a\ast}(A^1A^2A^0)$ and $\delta P_{\ast b}(B^1B^2B^0)$  in each of the groups on \emph{ibm\_kingston}. Notation as in Fig. \ref{belo}.}
\label{ss1k}
\end{figure*}

We have run the experiment in April 2025 on  \emph{ibm\_brisbane}, \emph{ibm\_sherbrooke} (Eagle generation), on the same groups of qubits,
and \emph{ibm\_torino}, \emph{ibm\_kingston} (Heron generation, revision 1 and 2, respectively),
with $60$ jobs, $20000$ shots, and $25$ repetitions of each set of settings $ab$, randomly shuffled,
so the total number of trials per setting is
\be
N=3\cdot 10^7=30000000=25\cdot 20000\cdot 60.
\ee
The single group is connected as in Fig. \ref{objsch} and the distribution of the tested 6 groups, see Tables \ref{ggg},\ref{gggt},\ref{gggk}, over the whole network is shown in Figs. \ref{objqub},\ref{objqubt},\ref{objqubk},
with indicated violation of Bell inequality and/or no-signaling.
The result of Bell expressions $\mathcal B$ (\ref{chshi}) is given in Table \ref{gggb}, \ref{gggbt} and Figs. \ref{belo}, while the test of no-signaling is
shown in Figs. \ref{ss1b},\ref{ss1s}, \ref{ss1t},\ref{ss1k}. The qubit frequencies for \emph{ibm\_brisbane} and \emph{ibm\_sherbrooke} are listed in Tables \ref{fffb}, \ref{fffs}, while the others are not disclosed.

The data show that the inequality has been violated by more than 10 standard deviations in the case of 3 groups \emph{ibm\_brisbane} and 5 groups \emph{ibm\_sherbrooke}.
Unfortunately, the violation is accompanied by a violation of no-signaling in part of the groups, especially on \emph{ibm\_sherbrooke}, which confirms the seriousness of this issue at IBM Quantum
\cite{nosig}. 
Even acknowledging that the signaling is much smaller than Bell violation and may be related to some crosstalk, we are unable to 
point out a direct cause. The common ZZ crosstalk, i.e., in the energy eigenspace, is unlikely as the choice operations $S_{\alpha/\beta}$ differ only by
phase, not amplitude. There is also no frequency collision \cite{well}. 
The Heron device, \emph{ibm\_torino} shows much smaller signaling. At the newest \emph{ibm\_kingston} all groups show violation of Bell inequality and
signaling stays within statistical error.

\section{Setup on IonQ}

\begin{figure*}
\includegraphics[scale=.5]{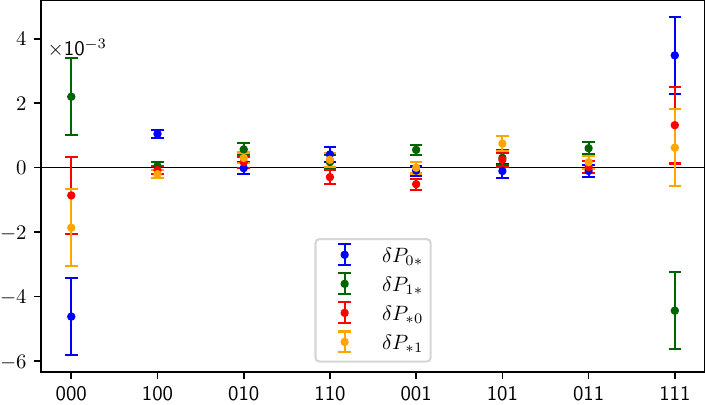}
\caption{Signaling between qubits $\delta P_{a\ast}(A^1A^2A^0)$ and $\delta P_{\ast b}(B^1B^2B^0)$  in each of the groups on IonQ Forte Enterprise 1. Notation as in Fig. \ref{belo}.}
\label{ss1i}
\end{figure*}

We were able to compare the IBM results with the competitive ion trap technology using IonQ quantum computer, Forte Enterprise 1, with cloud access.
It uses ytterbium ions ${}^{171}\mathrm{Yb}^+$ to support two-level qubit space, with the drive frequency 12.64GHz between hyperfine levels, placed in a Paul trap, see
details in the documentation \cite{iondoc,avsion,wright_phd,egan_phd,debnath_phd,landsman_phd,figgatt_phd}.
Due to all-to-all connectivity, the test requires only 6 qubits, and no additional separating one, with two entangled qubits and four friends. We have run each setting, randomly shuffled, with $10000$ shots
and $35$ repetitions. Unlike IBM, iterating over circuits and then shots, IonQ iterates over shots and then over circuits, which allows in principle allows memory effects between runs.
The timescales are much longer, with single qubit gate of $63\mu$s, two-qubit gate $650\mu$s, and readout $250\mu$s, compared to coherence times $\sim 100$s. Errors of single and two-qubit gates are
$5\cdot 10^{-4}$ and $6.6\cdot 10^{-3}$ respectively. The circuit is depicted in Fig. \ref{cirion},
and the details in Appendix \ref{appa}. The result of Bell combination is $2.569 \pm 0.0024$, which is undoubtedly violating the classical bound $2$.
There is also significant violation of no-signaling, Fig. \ref{ss1i}, but further tests are necessary to check if it is accidental.

The full data and scripts are publicly available \cite{zen}.

\section{No-signaling and the reliability of the Bell test}

In several cases, we observed violations of no-signaling. One could argue that it invalidates the violation of CHSH inequality.
Indeed, Bell theorem relies on the assumption of locality, and no-signaling is also its consequence. 
On the other hand, lack of signaling still does not verify the assumption, which is only falsifiable by its nature.
In early photonic experiments apparent signaling could appear due to postselections \cite{sig25}.
However, even new loophole-free tests \cite{nist,storz} show moderate signaling \cite{ab17,nosig}.
Instead of invalidating all tests, one can estimate a potential deviation from the locality assumption and how it changes the CHSH $\mathcal B$ value.
Note that relaxing the locality does not necessarily lead to signaling. For instance, there exists a simple model of nonlocal hidden variables $A_{xy}$ and $B_{xy}$
of the outcomes depending on the $xy$ choices of $AB$ parties, respectively. Namely, assigning
\be
A_{xy}=A_{00},\:B_{xy}=(-1)^{x+y+xy}A_{00}
\ee
with a random distribution of $A_{00}=\pm 1$ ($+1$ and $-1$ with probability $1/2$) gives $\mathcal B=4$, while $p(A\ast)=p(\ast B)=1/2$ regardless of the setting,
so no-signaling holds.

Nevertheless, one can believe that the magnitude of signaling should be comparable to the change of $\mathcal B$.
In our collected data $\mathcal B-2$ is in the range from $0.03$ to $0.5$ while signaling is in the range from
$\sim 10^{-4}$ for $10^{-2}$. Moreover, the newest IBM device 
 \emph{ibm\_kingston} gives a higher $\mathcal B$ with no evidence of signaling.
 We emphasize again that even passing the no-signaling test is not a verification but the absence of falsification.
 Therefore further more precise Bell tests, with signaling checks, are very important.

\section{Discussion}

The Bell inequality strengthened by the objective friends is violated only in some cases at a quantum processor. The fact that only
some cases show the violation, which is anyway far below the quantum ideal value, indicates problematic error propagation.
It calls for further research in order to identify the origins and countermeasures. It seems that the newest generation of IBM Heron, revision 2,
represented by \emph{ibm\_kingston} has overcome some problems of earlier devices, with all groups passing the test.
Anyway, successful violations give confidence in entanglement as
a nonclassical feature and sufficient device quality to spread entanglement over medium-range distances. 
Further improved tests should set the limit on distances and the number of friends able to violate the inequality.
One can combine it with tests of local friendliness \cite{wigf1, friendslo2}, dropping or weakening the assumption of absoluteness since sharing the same value is verifiable by simultaneous measurement by different friends.
Of serious concern is the violation of no-signaling, which indicates at least serious technical failures.
Resolving such problems is necessary for progress in reliable quantum computation.

\acknowledgements

JKK acknowledges the financial support of the
National Science Centre (NCN) through the QuantEra
project Qucabose 2023/05/Y/ST2/00139. We acknowledge
the use of IBM Quantum services for experiments in this
paper. The views expressed are those of the authors, and
do not reflect the official policy or position of IBM or the
IBM Quantum team.

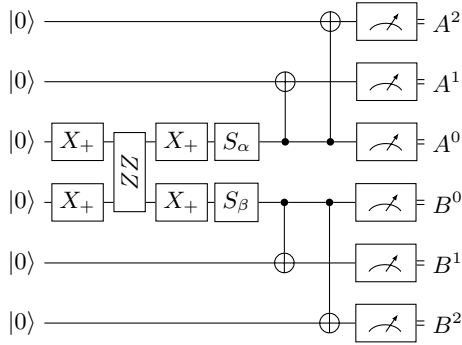
\begin{figure}
\begin{tikzpicture}[scale=1]
		\begin{yquant*}[register/separation=3mm]
			init { $\ket 0$} q[0,1,2,3,4,5];
			box {$X_+$} q[2];
			box {$X_+$} q[3];
			box {\rotatebox{90}{$ZZ$}} (q[2,3]);
			box {$X_+$} q[2];
			box {$X_+$} q[3];
			align -;
			box {$S_\alpha$} q[2];
			box {$S_\beta$} q[3];
			cnot q[1] | q[2];
			cnot q[0] | q[2];
			cnot q[4] | q[3];
			cnot q[5] | q[3];
			
			measure q[0-2];
			measure q[3-5];
			output {$A^2$} q[0];
			output {$A^1$} q[1];
			output {$A^0$} q[2];
			output {$B^2$} q[5];
			output {$B^1$} q[4];
			output {$B^0$} q[3];
		\end{yquant*}
\end{tikzpicture}
\caption{The circuit of the Bell test with friends on IonQ. The operational flow is from left to right.}\label{cirion}
\end{figure}

\begin{table}
\begin{tabular}{*{8}{c}}
\toprule
G&$A^2$&$A^1$&$A^0$&$M$&$B^0$&$B^1$&$B^2$\\
\midrule
1&3&5&4&15&22&21&23\\
2&27&29&28&35&47&46&48\\
3&40&42&41&53&60&59&61\\
4&65&67&66&73&85&84&88\\
5&78&80&79&91&98&97&99\\
6&103&105&104&111&122&121&123\\
\bottomrule
\end{tabular}
\caption{Groups of qubits used in the demonstration on \emph{ibm\_brisbane} and \emph{ibm\_sherbrooke}}
\label{ggg}
\end{table}

\begin{table}
\begin{tabular}{*{8}{c}}
\toprule
G&$A^2$&$A^1$&$A^0$&$M$&$B^0$&$B^1$&$B^2$\\
\midrule
1&3&5&4&16&23&22&24\\
2&28&30&29&36&48&47&49\\
3&41&43&42&54&61&60&62\\
4&66&68&67&74&86&85&87\\
5&79&81&80&92&99&98&100\\
6&104&106&105&112&124&123&125\\
\bottomrule
\end{tabular}
\caption{Groups of qubits used in the demonstration on \emph{ibm\_torino}}
\label{gggt}
\end{table}

\begin{table}
\begin{tabular}{*{8}{c}}
\toprule
G&$A^2$&$A^1$&$A^0$&$M$&$B^0$&$B^1$&$B^2$\\
\midrule
1&2&4&3&16&23&22&24\\
2&28&30&29&38&49&48&50\\
3&42&44&43&56&63&62&64\\
4&68&70&69&78&89&88&90\\
5&82&84&83&96&103&102&104\\
6&108&110&109&118&129&128&130\\
\bottomrule
\end{tabular}
\caption{Groups of qubits used in the demonstration on \emph{ibm\_kingston}}
\label{gggk}
\end{table}

\begin{table}
\begin{tabular}{*{8}{c}}
\toprule
G&$\mathcal B_b$&$\Delta \mathcal B_b[10^{-4}]$&$A-B$&$\mathcal B_s$ &$\Delta \mathcal B_s[10^{-4}]$&$A-B$\\
\midrule
1&2.259&2.75&$\leftarrow$&1.940&2.62&$\leftarrow$\\
2&1.843&2.69&$\leftarrow$&2.099&2.73&$\leftrightarrow$\\
3&1.981&2.74&$\leftrightarrow$&2.176&2.62&$\leftrightarrow$\\
4&2.034&2.67&&2.153&2.73&\\
5&1.872&2.94&&2.191&2.67&$\leftrightarrow$\\
6&2.194&2.79&&2.212&2.81&$\leftrightarrow$\\
\bottomrule
\end{tabular}
\caption{Results of Bell tests $\mathcal B_{b/s}$ (\ref{chshi}) on each group with the error  $\Delta \mathcal B_{b/s}$ for \emph{ibm\_brisbane}/\emph{ibm\_sherbrooke}, respectively, for each group of qubits.
The significant violation of no signaling is marked by $A\to B$, $A\leftarrow B$, $A\leftrightarrow B$, for the respective direction of signaling.
The criterion of significance is the nonzero value beyond 5 standard deviations}
\label{gggb}
\end{table}

\begin{table}
\begin{tabular}{*{8}{c}}
\toprule
G&$\mathcal B_t$&$\Delta \mathcal B_t[10^{-4}]$&$A-B$&$\mathcal B_k$ &$\Delta \mathcal B_k[10^{-4}]$&$A-B$\\
\midrule
1&1.959&2.66&&2.535&2.62&\\
2&2.122&2.74&&2.045&2.82&\\
3&2.000&2.62&&2.478&2.68&\\
4&1.493&2.64&&2.379&2.72&\\
5&2.040&2.68&$\rightarrow$&2.443&2.67&\\
6&1.551&2.74&&2.444&2.63&\\
\bottomrule
\end{tabular}
\caption{Results of Bell tests $\mathcal B_{t/k}$ (\ref{chshi}) on each group with the error  $\Delta \mathcal B_{t/k}$ for \emph{ibm\_torino}/\emph{ibm\_kingston}, respectively, for each group of qubits.
The significant violation of no signaling is marked by $A\to B$, $A\leftarrow B$, $A\leftrightarrow B$, for the respective direction of signaling.
The criterion of significance is the nonzero value beyond 5 standard deviations}
\label{gggbt}
\end{table}

\begin{table}
\begin{tabular}{*{8}{c}}
\toprule
G&$A_2$&$A_1$&$A_0$&$M$&$B_0$&$B_1$&$B_2$\\
\midrule
1&4875&4734&4818&4947&5038&4966&4844\\
2&4747&4628&4994&4907&4769&4853&4844\\
3&4868&5063&4935&4996&4766&4971&4793\\
4&4957&5113&4885&4979&5104&4679&4898\\
5&4643&5031&4863&4914&4942&4816&5060\\
6&4933&4982&4765&4853&4937&4967&5034\\
\bottomrule
\end{tabular}
\caption{Drive frequencies [MHz] of qubits used in the demonstration on \emph{ibm\_brisbane}}
\label{fffb}
\end{table}

\begin{table}
\begin{tabular}{*{8}{c}}
\toprule
G&$A_2$&$A_1$&$A_0$&$M$&$B_0$&$B_1$&$B_2$\\
\midrule
1&4747&4850&4787&4599&4668&4773&4758\\
2&4680&4791&4742&4914&4795&4674&4707\\
3&4705&4654&4814&4767&4672&4810&4901\\
4&4758&4893&4809&4874&4745&4666&4889\\
5&4861&5042&4786&4893&4774&4949&4832\\
6&4694&4883&4783&4899&4731&4847&4820\\
\bottomrule
\end{tabular}
\caption{Drive frequencies [MHz] of qubits used in the demonstration on \emph{ibm\_sherbrooke}}
\label{fffs}
\end{table}

\appendix
\section{Implementation of necessary gates}
\label{appa}

\begin{figure}
\begin{tikzpicture}[scale=1]
		\begin{yquantgroup}
			\registers{
			qubit {} q[2];
			}
			\circuit{
			init {$F$} q[0];
			init {$G$} q[1];
			box {$\downarrow$}  (q[0,1]);}
			\equals
			\circuit{
			box {\rotatebox{90}{$CR^+$}}  (q[0,1]);
			box {$X$} q[0];
			box {\rotatebox{90}{$CR^-$}}   (q[0,1]);}
		\end{yquantgroup}
\end{tikzpicture}

\caption{The notation of the ECR gate in the convention $ECR_\downarrow|FG\rangle$}
\label{ecr}
\end{figure}

To run our test, we require $CX$ gate,
\be
CX_\downarrow
=\begin{pmatrix}
1&0&0&0\\
0&1&0&0\\
0&0&0&1\\
0&0&1&0\end{pmatrix}
\ee
in the basis
$|00\rangle$, $|01\rangle$, $|10\rangle$, $|11\rangle$.
We denote reversed gates by $\langle F'G'|U_\uparrow|FG\rangle=\langle G'F'|U_\downarrow |GF\rangle$.
The ideal $CX$ gate is its own reverse, real, and Hermitian. 
The $CX$ gates are actually transpiled by echo cross-resonance gates ($ECR$);  see below.

For the physical implementation and manipulation of qubits as transmons \cite{transmon} and programming of IBM Quantum, see \cite{gam0,qurev,qisr}.
The IBM Quantum devices use a native asymmetric two-qubit $ECR$ gates \cite{ecr1,ecr2} instead of $CX$, but one can transpile the latter by the former, adding single-qubit gates.
We shall use Pauli matrices in the basis $|0\rangle$, $|1\rangle$,
\be
X=\begin{pmatrix}
0&1\\
1&0\end{pmatrix},\:Y=\begin{pmatrix}
0&-i\\
i&0\end{pmatrix},\:Z=\begin{pmatrix}
1&0\\
0&-1\end{pmatrix},\:
I=\begin{pmatrix}
1&0\\
0&1\end{pmatrix}.\label{pauli}
\ee

The $ECR$ gate acts on the states $|FG\rangle$ as (Fig. \ref{ecr})
\ba
&ECR_\downarrow=(XI-YX)/\sqrt{2}=CR^- (XI) CR^+=\nonumber\\
&
\begin{pmatrix}
0&X_-\\
X_+&0\end{pmatrix}
=\begin{pmatrix}
0&0&1&i\\
0&0&i&1\\
1&-i&0&0\\
-i&1&0&0\end{pmatrix}/\sqrt{2},
\ea
in the basis $|00\rangle$, $|01\rangle$, $|10\rangle$, $|11\rangle$
where the native gate is
\be
X_+=X_{\pi/2}=(I-iX)/\sqrt{2}=\begin{pmatrix}
1&-i\\
-i&1\end{pmatrix}/\sqrt{2},
\ee
and $X_-=X_{-\pi/2}=ZX_+Z$, 
with 
\be
CR^\pm=(ZX)_{\pm \pi/4},
\ee
using the convention  $V_\theta=\exp(-i\theta V/2)=\cos(\theta/2)-iV\sin(\theta/2)$ if $V^2=I$ or $II$.
The gate is its inverse, i.e. $ECR_\downarrow ECR_\downarrow=II$.

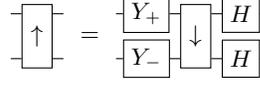
\begin{figure}[H]
\begin{tikzpicture}[scale=1]
		\begin{yquantgroup}
			\registers{
			qubit {} q[2];
			}
			\circuit{
			box {$\uparrow$} (q[0,1]);
			}
			\equals
			\circuit{
			box {$Y_+$}  q[0];
			box {$Y_-$}   q[1];
			box {$\downarrow$}  (q[0,1]);
			box {$H$}  q[0];
			box {$H$}   q[1];
			}
		\end{yquantgroup}
\end{tikzpicture}

\caption{The $ECR_\uparrow$ gate expressed by $ECR_\downarrow$ }
\label{ecrr}
\end{figure}

Note that $Z_\theta=\exp(-i\theta Z/2)=\mathrm{diag}(e^{-i\theta/2},e^{i\theta/2})$ is a virtual gate adding essentially the phase  shift to the next gates.
The actually performed operations always have the form $Z_{-\theta}UZ_\theta$, where $U$ is some basis gate.
 $ECR$ is asymmetric, i.e., the qubits are not interchangeable, but it can be reversed, i.e., for $F\leftrightarrow G$, (Fig. \ref{ecrr})
\be
ECR_\uparrow=(IX-XY)/\sqrt{2}=(HH) ECR_\downarrow(Y_+Y_-),
\ee
denoting $V_\pm= V_{\pm \pi/2}$, and Hadamard gate,
\be
H=(Z+X)/\sqrt{2}
=Z_+X_+Z_+=\begin{pmatrix}
1&1\\
1&-1\end{pmatrix}/\sqrt{2},
\ee
and $Z_\pm X_+Z_\mp=Y_\pm$, with $Y_+=HZ$ and $Y_-=ZH$.

\begin{figure}[H]
\begin{tikzpicture}[scale=1]
		\begin{yquantgroup}
			\registers{
			qubit {} q[2];
			}
			\circuit{
			box {$X$}  q[0];
			box {$X_+$}   q[1];
			box {$\downarrow$}  (q[0,1]);
			box {$Z_+$}  q[0];
			}
			\equals
			\circuit{
			cnot q[1] | q[0];
			}
		\end{yquantgroup}
\end{tikzpicture}

\caption{The $CX_\downarrow$ gate expressed by $ECR_\downarrow$ }
\label{cnot}
\end{figure}
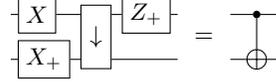

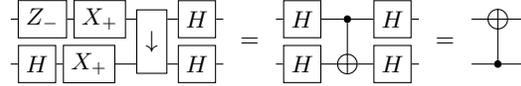
\begin{figure}[H]
\begin{tikzpicture}[scale=1]
		\begin{yquantgroup}
			\registers{
			qubit {} q[2];
			}
			\circuit{
			box {$Z_-$}  q[0];
			box {$H$}   q[1];
			box {$X_+$}  q[0];
			box {$X_+$}   q[1];
			box {$\downarrow$}  (q[0,1]);
			box {$H$}  q[0];
			box {$H$}  q[1];
			}
			\equals
			\circuit{
			box {$H$}  q[0];
			box {$H$}   q[1];
			cnot q[1] | q[0];
			box {$H$}  q[0];
			box {$H$}   q[1];
			}
			\equals
			\circuit{
			cnot q[0] | q[1];
			}
		\end{yquantgroup}
\end{tikzpicture}

\caption{The $CX_\uparrow$ gate expressed by $ECR_\downarrow$ }
\label{cnotr}
\end{figure}

The $CNOT$ gate can be expressed by $ECR$ (Fig. \ref{cnot}),
\ba
&CX_\downarrow=(II+ZI+IX-ZX)/2=\nonumber\\
&
\begin{pmatrix}
I&0\\
0&X\end{pmatrix}
=\begin{pmatrix}
1&0&0&0\\
0&1&0&0\\
0&0&0&1\\
0&0&1&0\end{pmatrix}\nonumber\\
&
=(Z_+ I)ECR_\downarrow (XX_+),
\ea
while its reverse reads (Fig. \ref{cnotr})
\ba
&CX_\uparrow=(II+IZ+XI-XZ)/2
=\nonumber\\
&\begin{pmatrix}
1&0&0&0\\
0&0&0&1\\
0&0&1&0\\
0&1&0&0\end{pmatrix}=(HH)X_\downarrow(HH)\nonumber\\
&
=
(HH)ECR_\downarrow (X_+X_+)(Z_-H).
\ea

IBM Heron devices have native $CZ$ gates, i.e. $CZ|ab\rangle=(-1)^{ab}|ab\rangle$ (only flips the sign of the $|11\rangle$ state)
and $CX_\downarrow=(IH)CZ(IH)$.

IonQ Forte native entangling gate is $(ZZ)_{\pi/2}=\exp(-iZZ\pi/4)$ that reads
\be
\begin{pmatrix}
1&0&0&0\\
0&i&0&0\\
0&0&i&0\\
0&0&0&1
\end{pmatrix}
\ee
which is symmetric and can express $CX$ gate as
\be
(IY_+)(ZZ)_{\pi/2}(IX_-)(Z_-Z_-)
\ee
or equivalently, see Fig. \ref{cnotion}.

\begin{figure}[H]
\begin{tikzpicture}[scale=1]
		\begin{yquantgroup}
			\registers{
			qubit {} q[2];
			}
			\circuit{
			box {$Z_-$} q[1];
			box {$Z_-$}  q[0];
			box {$X_-$}  q[1];
			box {\rotatebox{90}{$ZZ$}}   (q[0,1]);
			box {$Y_+$}  q[1];	
			}
			\equals
			\circuit{
			cnot q[1] | q[0];
			}
		\end{yquantgroup}
\end{tikzpicture}

\caption{The $CX_\downarrow$ gate expressed by $ZZ$ gates on IonQ. The two 
$Z_-$ rotations are omitted in the actual runs: the target friend qubits start in 
$|0\rangle$ and the control-side rotation commutes through the subsequent copying gates into the
$Z$-basis readout.
}
\label{cnotion}
\end{figure}

\end{document}